\documentclass[aps,prb,amsmath,amssymb,reprint,author-year,author-numerical,floatfix,superscriptaddress]{revtex4-2}
\usepackage{graphicx}
\graphicspath{{./images/}}
\usepackage{dcolumn}
\usepackage{bm}
\usepackage{mathtools}
\usepackage{amssymb}
\expandafter\let\csname equation*\endcsname\relax
\expandafter\let\csname endequation*\endcsname\relax
\usepackage{amsmath}
\usepackage{amsbsy}
\usepackage{booktabs}
\usepackage{verbatim}
\usepackage{array}
\usepackage{braket}
\newcolumntype{P}[1]{>{\centering\arraybackslash}p{#1}}
\usepackage{xcolor}
\usepackage[colorlinks = true,
            linkcolor = blue,
            urlcolor  = blue,
            citecolor = blue,
            anchorcolor = blue]{hyperref}
\AtBeginDocument{}%

\begin{document}

\preprint{AIP/123-QED}

\title[FISCHETTI]{On the static dielectric constant of thin dielectrics in extremely scaled silicon nanosheet transistors}

\author{Massimo V. Fischetti}
\email[email: ]{max.fischetti@utdallas.edu.}
\affiliation{Department of Materials Science and Engineering, The University of Texas at Dallas\\
             800 W. Campbell Rd., Richardson, TX 75080}   
\author{Dallin O. Nielsen}
\affiliation{Department of Materials Science and Engineering, The University of Texas at Dallas\\
             800 W. Campbell Rd., Richardson, TX 75080}
\author{Edward Chen}
\affiliation{Corporate Research, Taiwan Semiconductor Manufacturing Company Ltd.\\
             8 168, Park Ave. II, Hsinchu Science Park, Hsinchu 300-75, Taiwan}                         

\date{\today}

\begin{abstract}
We argue that the static dielectric constant of small (thin and/or narrow) semiconductor and insulator nanostructures depends strongly on the their
environment. We do so by considering the electronic response simply reviewing, briefly but critically, the existing literature. Regarding the ionic
response, in addition to reviewing the literature, we use a simple model to account for the confinement of optical phonons in thin films and show that the
reduction of their density of states has a negligible effect on the dielectric constant, in contrast to some claims found in the literature. In general,
we argue that in realistic structures, such as double-gated Si nanosheets, the use of the bulk dielectric constants for both the channel and the gate insulators, is justified.
\end{abstract}

\maketitle

\section{\label{sec:level1}{Introduction}} 
The continuing scaling of semiconductor devices requires the use of thinner semiconductor and insulating films. For example, the field effect 
transistors (FETs) of the present 2--3~nm node employ channels consisting of Si nanosheets as thin as 3 unit cells ($\approx$~1.5~nm)~\cite{Agrawal_2024} and (double) gate-insulator stacks consisting of a 0.6~nm-thick SiO$_2$ layer and a 1.2~nm-thick HfO$_2$ 
film~\cite{Loubet_2017,Park_2022,Yeap_2024,Agrawal_2024}. In the study of the electrostatic and charge-transport properties of these devices the 
question arises of whether the dieectric properties of these films (specifically, their static dielectric constant) are affected by their reduced thickness.
Obviously, quantum confinement changes the electronic band structure as electrons must be considered as a two-dimensional electron gas (2DEG). However, also their dielectric properties change when considering very thin films. Since the dielectric response of a crystal is due to its electronic and ionic polarization,
in thin films we may expect a different electronic response due to the change of the bandgap, usually increasing in thinner films, and a change of the ionic
response due the different density of states of the optical phonons that may be geometrically confined.\\ 

Here we intend to discuss these issues in general terms, considering the covalent Si and polar SiO$_2$ and HfO$_2$ as specific examples, dealing separately
with these two types of polarization: the electronic polarizability (that controls entirely the dielectric response of covalent crystals, such as indeed Si) and the ionic response that plays a large role in controlling the dielectric constant of polar insulators. To anticipate and summarize the main conclusion 
from a survey of the literature, looking at both experimental and theoretical work, the static dielectric constant of thin films of semiconductors or insulators decreases with decreasing thickness when considering isolated, free-standing films. However, in films embedded in other
solids, such as Si supported and/or gated by SiO$_2$ or HfO$_2$, the dielectric constant changes only by a small amount. Interface effects, such as surface 
polarization or the presence of interfacial layers, may even increase it for both semiconductors and insulators: In short, the polarizability of a free surface 
is reduced by the proximity of the vacuum, but it may even be boosted by the proximity of a medium with a higher polarizability. Overall, we do not think that 
it is a major effect.\\  

\begin{figure}[tb]
\centerline 
{\vbox{
{\hbox{
\includegraphics[width=6.60cm]{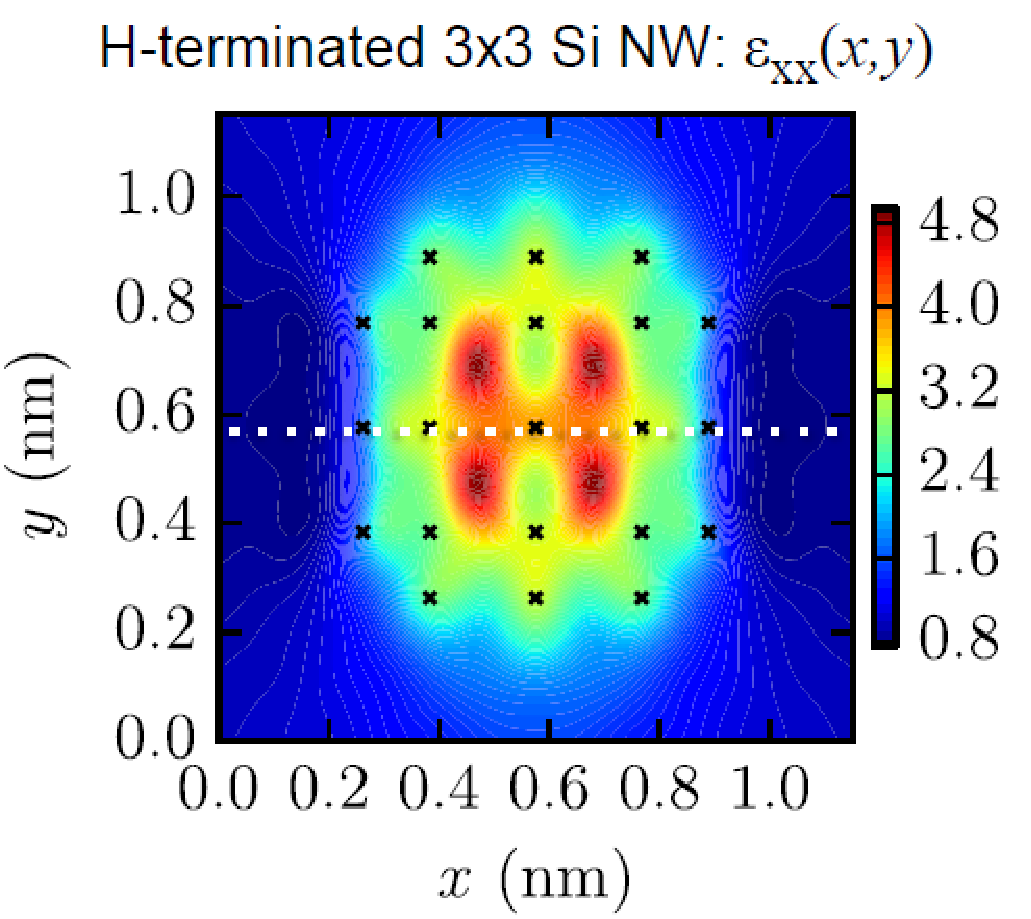}}}
{\hbox{
\includegraphics[width=6.6cm]{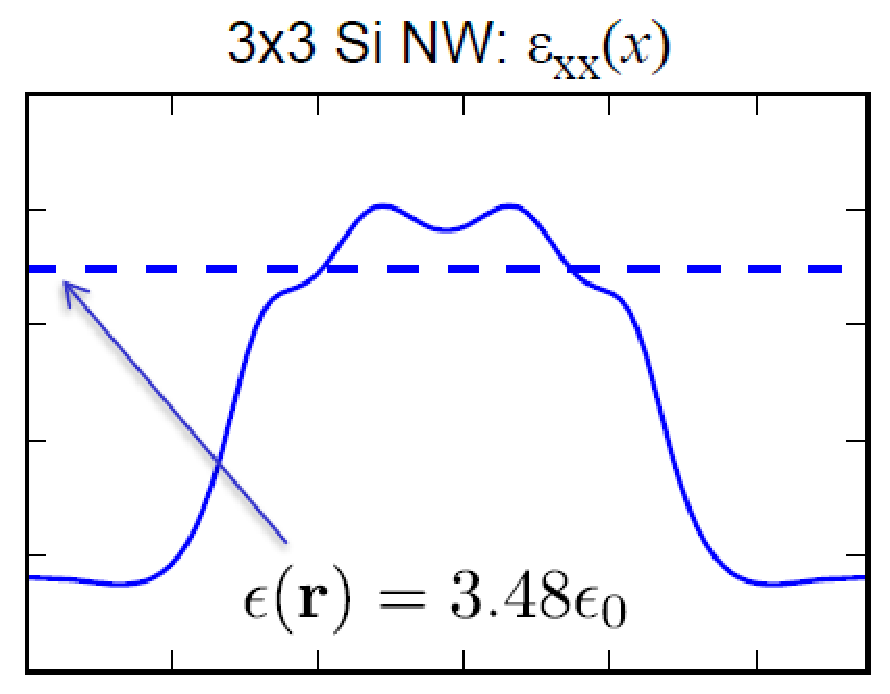}}}
}}
\caption{{\bf Top}: Contour plot of the local dielectric constant of a freestanding 3x3 cells Si nanowire calculated using the method described in
        Ref.~\cite{Fang_2016}. {\bf Bottom}: The same but along the cut shown by the dashed line at left. Note how the dielectric constant is reduced
        significantly from its bulk value. Presumably, this is due to the fact that the nanowire is surrounded by vacuum. The presence of a high-$\kappa$
        dielectric is likely to increase the polarizability of the nanowire, thus increasing its dielectric constant, as one may infer from the
        literature. [Unpublished work by Jingtian Fang.]}  
\label{fig:SiNW_epsilon}
\end{figure}
\section{Electronic response}
\begin{figure*}[tb]
\centerline
{\hbox{
\includegraphics[width=6.50cm]{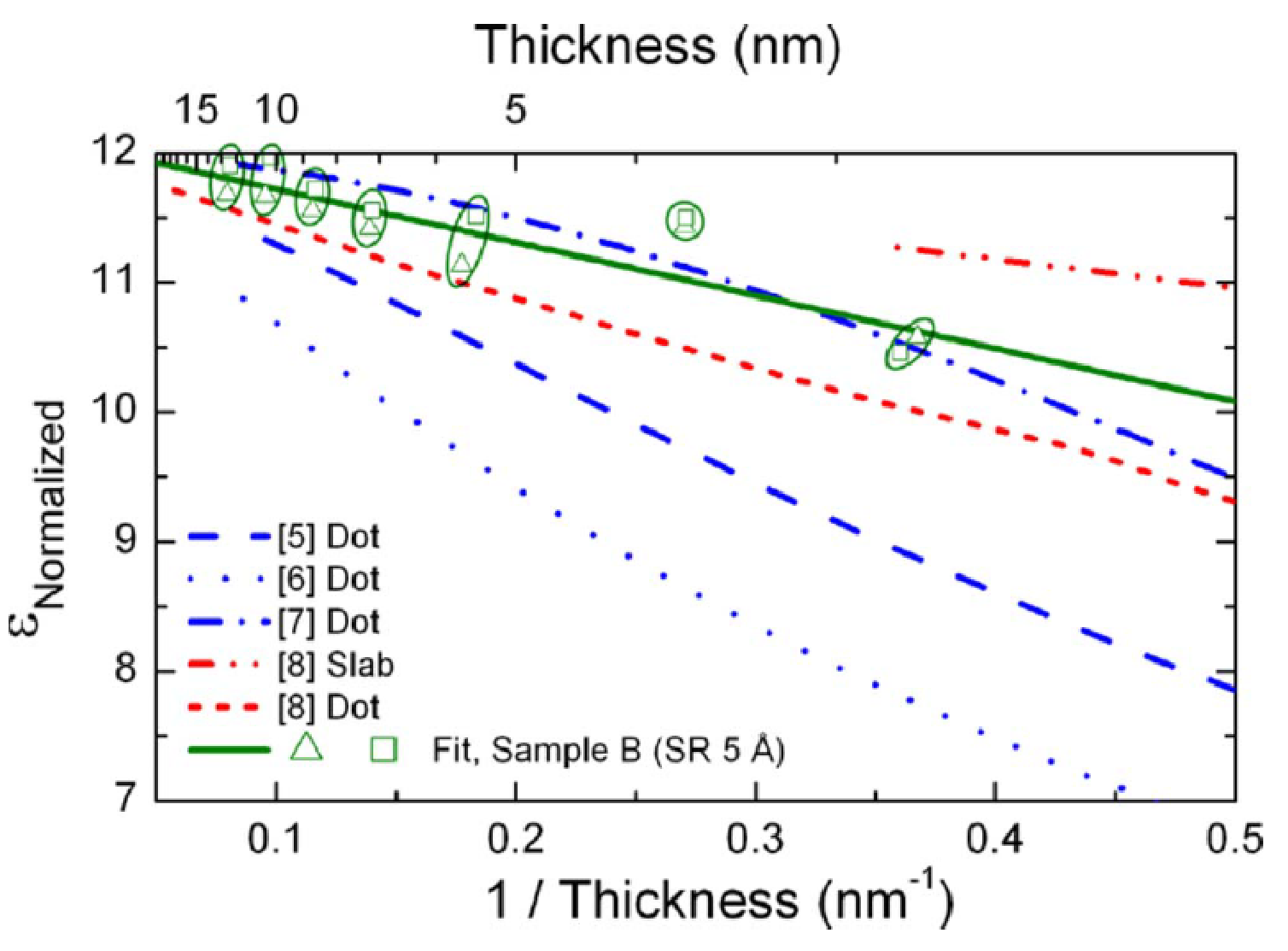}
\includegraphics[width=5.70cm]{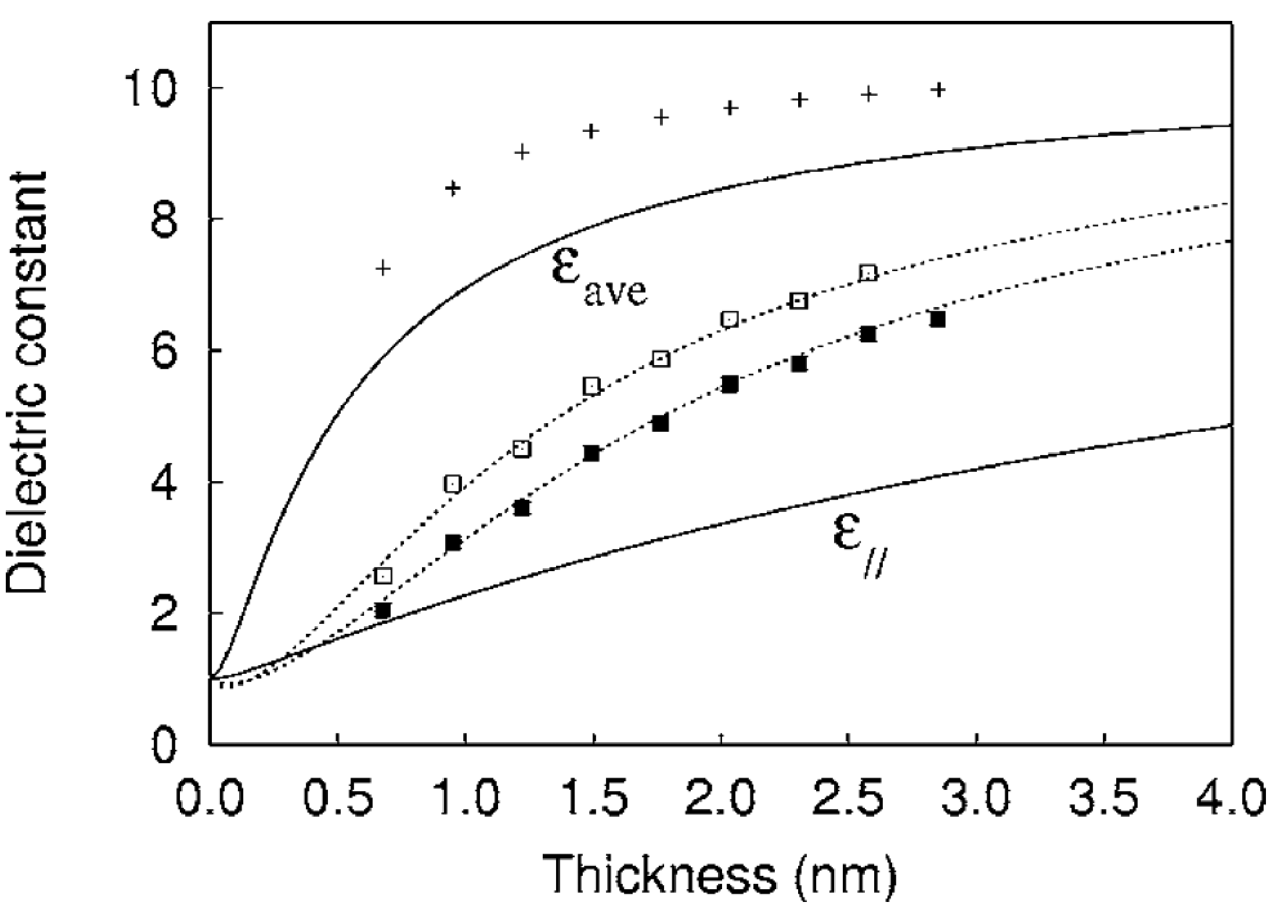}
\includegraphics[width=5.60cm]{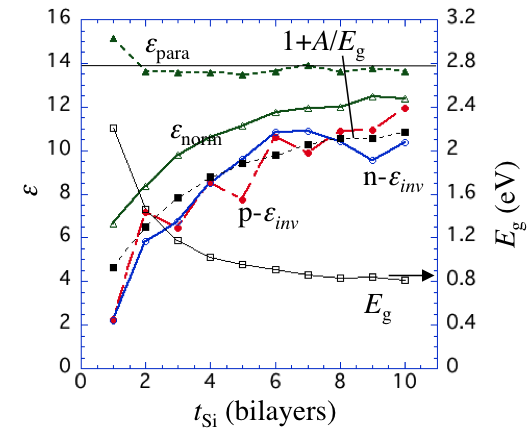}
}}
\caption{{\bf Left}: Static dielectric constant of thin free-standing Si films plotted as a function of film 
        thickness. The experimental data, obtained from ellipsometry, are shown by the open green symbols and the thick green line. Results from various 
        theoretical models for Si quantum dots and thin films are also shown. [Reprinted with permission from H.~G.~Yoo and P.~Fauchet, Phys. Rev. 
        B {\bf 77}, 115355 (2008) \copyright2008 by the American Physical Society.]
        {\bf Center}: ``{\it Effective dielectric constant $\epsilon_{\rm M}$ of Si quantum wells (+) and of layers of cubic Si nanocrystals separated by a 
        vacuum gap of 2.7~$\AA$ (open squares) and 4.1~$\AA$ (solid squares) calculated in tight binding. The dotted line is the fit of these data using
        Eq~(1). The continuous lines represent the average dielectric constant $\epsilon_{\rm ave}$ and the dielectric constant $\epsilon_{\parallel}$ 
        of capacitors in parallel}''. [Figure and words in quotes/italic reprinted with permission from C.~Delerue and G.~Allan, Appl. Phys. Lett. 
        {\bf 88}, 173117 (2006) \copyright2006 by the American Physical Society.]
        {\bf Right}: ``{\it Channel thickness $t_{\rm Si}$ dependence of various dielectric constants. p-$\epsilon_{\rm inv}$ and n-$\epsilon_{\rm inv}$ are the 
        inversion-layer dielectric constants $\epsilon_{\rm inv}$ for the p- and the n-channel. All results are at $V_{\rm G}=\pm$~3~V. 
        $\epsilon_{\rm para}$ and $\epsilon_{\rm norm}$ are the static electric parts of the optical dielectric constants $\epsilon_{\rm opt}$ for
        incident light with different polarization. Also shown, the channel thickness $t_{\rm Si}$ dependence of the band gap $E_{\rm g}$}''.
        [Figure and words in quotes/italic reprinted/adapted with permission from H.~Kageshima and A. Fujiwara, Appl. Phys. Lett. {\bf 96}, 193102 (2010), 
        \copyright2010 by the American Institute of Physics.]}        
\label{fig:SiNW_epsilon_ref}
\end{figure*}

The electronic dielectric properties of thin Si films have been investigated both experimentally and theoretically.   
Experimentally, Yoo and Fauchet~\cite{Yoo_2008} using ellipsometry have observed a reduction from its bulk value of the dielectric constant of 
thin (100) Si films on SiO$_2$ from $\approx 11.9~\epsilon_{\rm vac}$ (where $\epsilon_{\rm vac}$ is the permittivity of vacuum)) in bulk Si to about
10.2~$\epsilon_{\rm vac}$ in 2~nm-thin Si films. Theoretically, the study is quite complicated and indeed the literature shows inconsistent results:
Using DFT in the local density approximation, Shi and Ramprasad~\cite{Shi_2007} have not predicted any depression of $\epsilon_{\rm Si}$ even 
in (111) Si films as thin as 1~nm. Specifically, they have employed Si-SiO$_2$ and Si-HfO$_2$ supercells (1.096~nm thick Si, 1.47~nm-thin SiO$_2$ in 
the $\beta$-cristobalite structure and and 1.99~nm-thin HfO$_2$ in the tetragonal structure), obtaining a static dielectric constant of about 
12~$\epsilon_{\rm vac}$ for Si and about 4~$\epsilon_{\rm vac}$ for SiO$_2$ and 16~$\epsilon_{\rm vac}$ for HfO$_2$ (whose bulk value in tetragonal 
is about the same). They also observed an increasing polarization of Si and oxides at the interfaces. Similar results have been obtained by 
Giustino {\it et al.}~\cite{Giustino_2003}, who have calculated a moderate reduction, about $8 \pm 1$\%, of $\epsilon_{\rm Si}$ is 1.5~nm Si layers 
surrounded by SiO$_2$. Whereas these results have been obtained considering Si films surrounded by dielectrics, a depression of $\epsilon_{\rm Si}$ 
has been predicted when considering free-standing Si sheets. For example, Kegeshima and Fujiwara~\cite{Kageshima_2008,Kageshima_2010} have used DFT 
with the generalized gradient approximation (GGA) to study H-terminated free-standing (111) Si films obtaining a significant depression of the Si 
dielectric constant, as large as a factor of 4, for the out-of-plane dielectric constant; specifically, the in-plane value of $\epsilon_{\rm Si}$ 
remains constant at around 13.5~$\epsilon_{\rm vac}$, whereas its out-of-plane value drops from 12.5~$\epsilon_{\rm vac}$ to 6.6~$\epsilon_{\rm vac}$
as the Si thickness is reduced from 10 bilayers to 1 bilayer. Similarly, Delerue and Allan~\cite{Delerue_2006} considered single layers of cubic and 
spherical Si nanocrystals using self-consistent tight binding and found that that $\epsilon_{\rm Si}$ is reduced from its bulk value and is strongly
influenced by the morphology of the nanostructure. Judging from these results, one may speculate that indeed the dielectric environment and the 
morphology of the interfaces affect significantly the dielectric response and that a thin film may inherit to some extent the polarization properties 
of the surrounding environment. This is confirmed by the fact that Ref.~\cite{Shi_2007} shows that thin Si films surrounded by SiO$_2$ and HfO$_2$ 
exhibit an ionic response, presumably induced by the surrounding oxides. Moreover, even Yoo and Fauchet~\cite{Yoo_2008} observe a significant effect of the
polarization of the surface; in their samples this is the Si-air interface and this may cause the moderate reduction of $\epsilon_{\rm Si}$ that they observe. Similarly, our own work (unpublished work for Si but published for graphene and graphene nanoribbon~\cite{Fang_2016}) based on the RPA within an empirical-pseudopotential approximation, shows a very strong reduction of the static, long-wavelength dielectric tensor in ultra-small free-standing (100) Si nanowires. 
This is shown in Fig.~\ref{fig:SiNW_epsilon}. The experimental results of Refs.~\cite{Yoo_2008}, \cite{Kageshima_2010}, and \cite{Delerue_2006} are also 
shown in Fig.~\ref{fig:SiNW_epsilon_ref}.\\

It may be tempting to view the increasing bandgap as the main (or only) cause of the reduction of the dielectric constant in thin films: Indeed, considering
the static, long-wavelength limit of the well-known Lindhart expression for the polarization of the valence electrons in a dielectric (see Eq.~(11.42) of
Ref.~\cite{thebook}), we have
\begin{equation}
\epsilon({\bf q}=0;\omega=0) \approx \epsilon_{\rm vac} \left [ 1+ \left ( \frac{\hbar \omega_{\rm v,P}}{E_{\rm gap}} \right )^2 \right ] \ ,
\label{eq:static_epsilon}
\end{equation}
where $\omega_{\rm v,P}$ is the frequency of the valence plasmons, and $E_{\rm gap}$ is the band gap. Therefore, we may expect that in thin films a larger
bandgap may reduce the dielectric constant approximately by the ratio $E^{\rm (bulk)}_{\rm gap}/E^{\rm (film)}_{\rm gap}$. 
The main complications is due to the fact that the bandgap $E_{\rm gap}$ appearing in this equation is the average of the direct gap throughout the
entire Brillouin zone, weighted by the joint valence-conduction density of states. In thin films, the bandgap may even decrease as the bulk Brillouin zone is 
projected onto the 2D plane. This is the case of Si that acquires a small direct bandgap at the symmetry point $\overline{\Gamma}$.  
Therefore, only detailed and accurate band-structure calculations of the RPA 
expression can provide an answer and the well-known problems that affect DFT calculations of the bandgap render the task quite hard. Moreover, in thin films,
quantum confinement may affect the result. We may also expect that the wavelength dependence of 2D plasmons, inter-subband plasma excitations, and surface
polarization may play a significant role.\\

In any event, the experimentally observed drop of the static dielectric constant is limited to less than 20\% in Si sheets as thin as 3 cells. Therefore, 
assuming the bulk Si value for the dielectric constant of thin Si nanosheets is not going to affect significantly the results of any calculation, as it is a variation smaller than the uncertainties due to the many other approximations that must be embraced in most cases.

\section{Ionic response} 
In insulators, the electronic response shows the same reduction expected for semiconductors. However, given their larger bandgap, quantum electronic confinement 
has a much smaller effect. 
Indeed, Ming {\it et al.}~\cite{Ming_2011} have measured for the bandgap of HfO$_2$ values increasing only by about 3\% as the thickness decreases from 4 to 2~nm
The issue is also complicated by the fact that HfO$_2$ can crystallize in various forms, monoclinic, tetragonal, orthorhombic, or cubic, each exhibiting a different
bandgap. This complicates the issue of understanding the behavior of this amorphous hafnia films.\\ 
\begin{figure*}[tb!]
\centerline
{\vbox{
{\hbox{
\includegraphics[width=8.50cm]{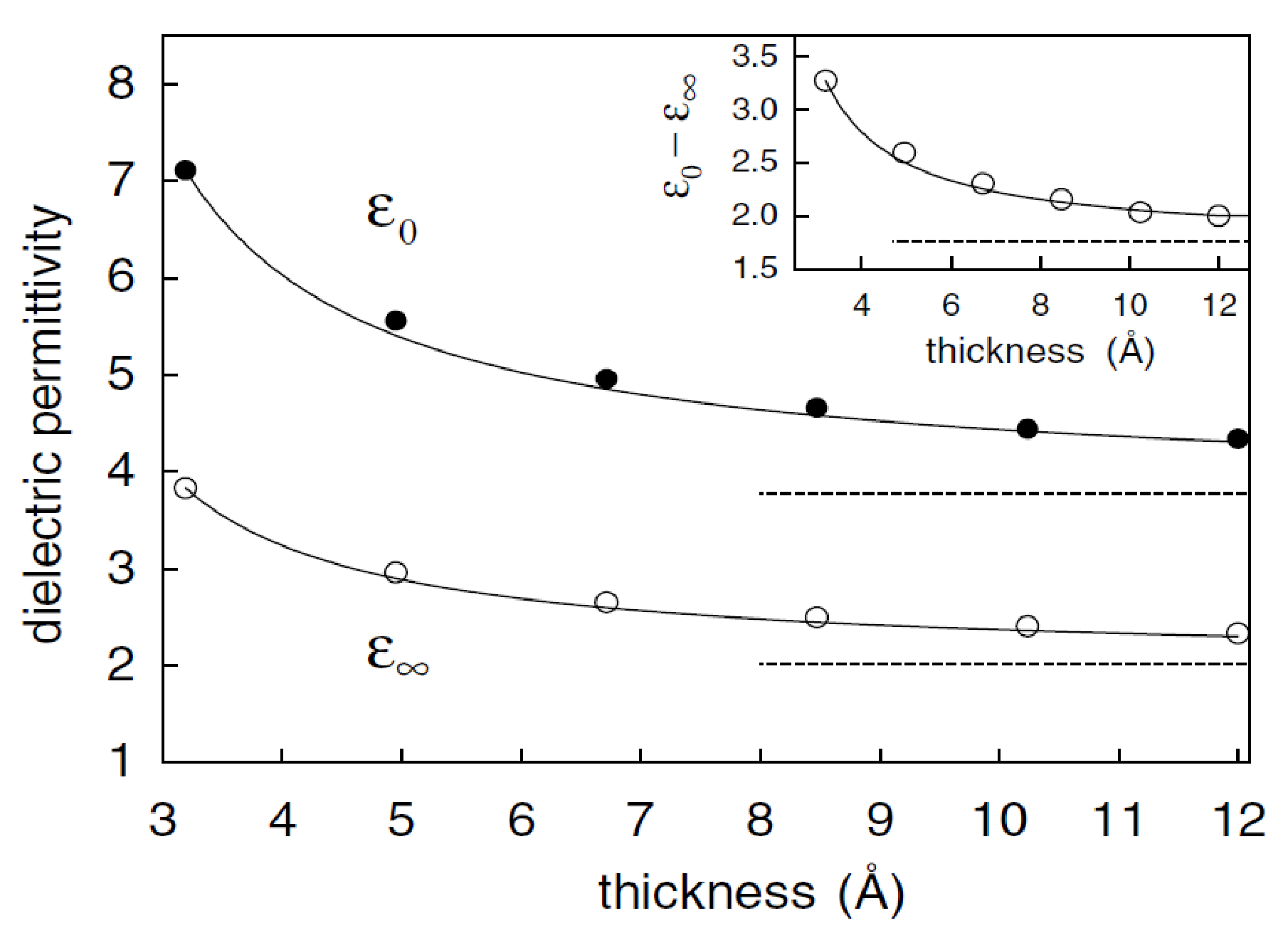}
\includegraphics[width=7.00cm]{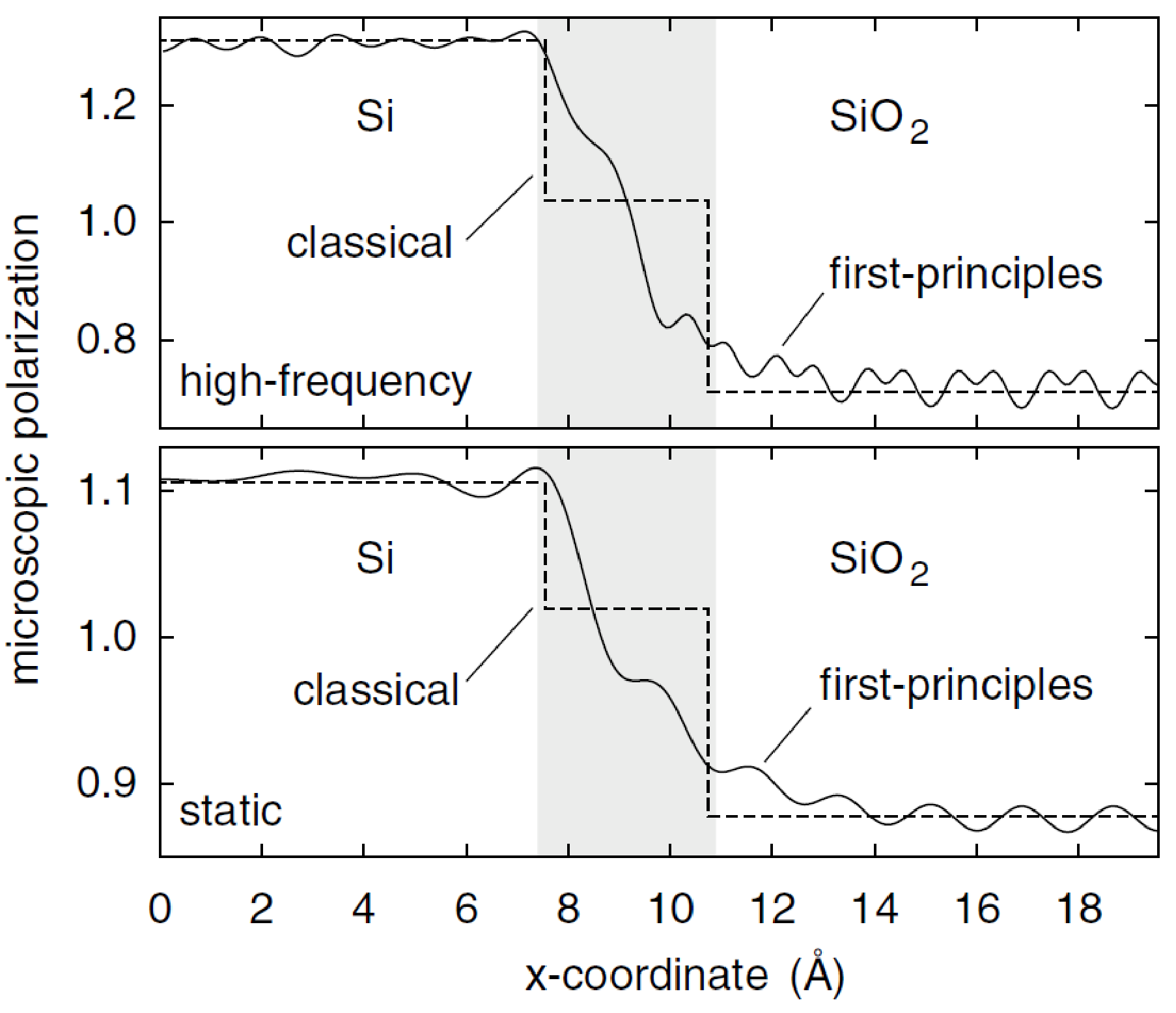}}}
{\hbox{
\includegraphics[width=8.0cm]{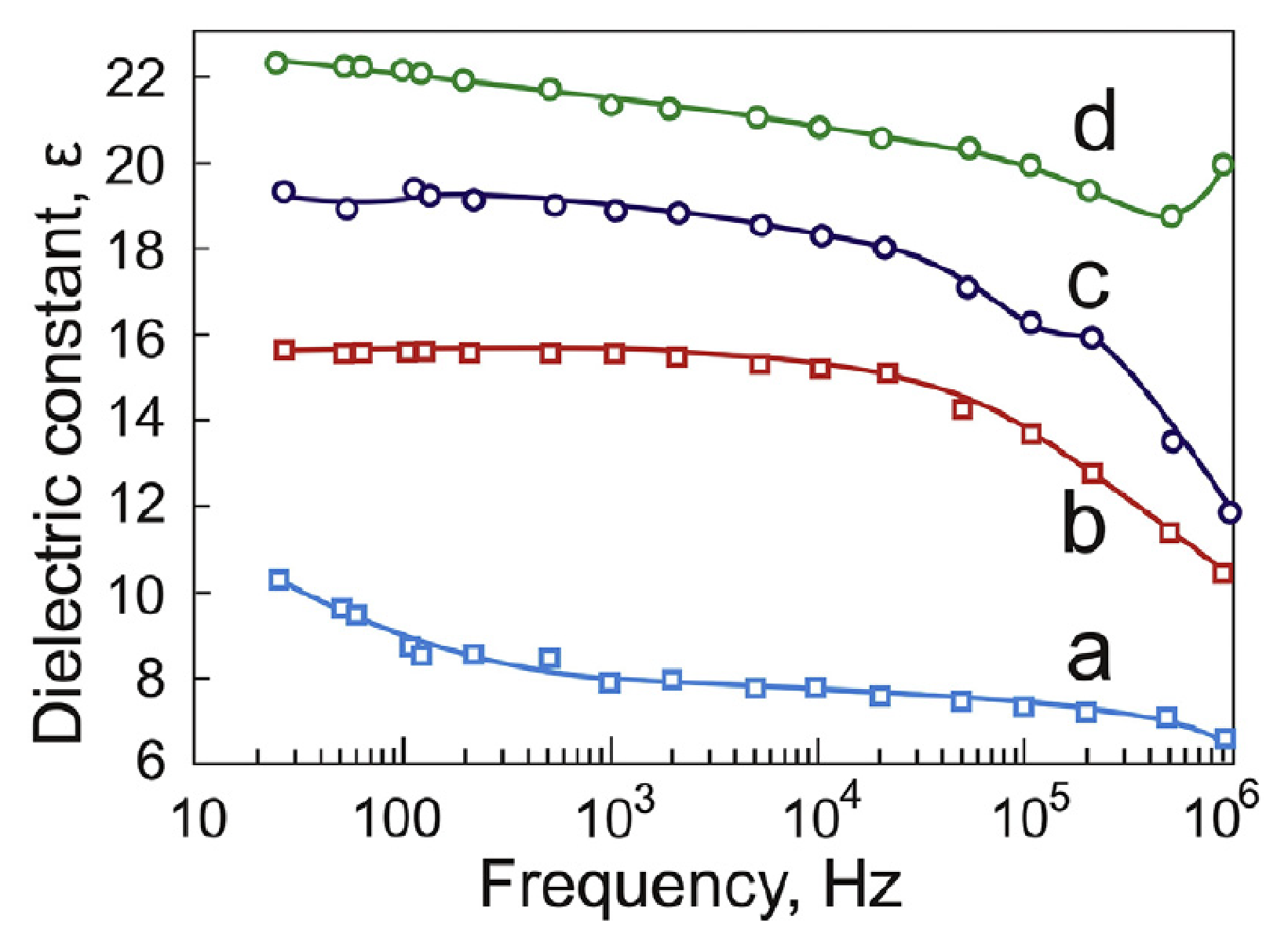}
\includegraphics[width=8.50cm]{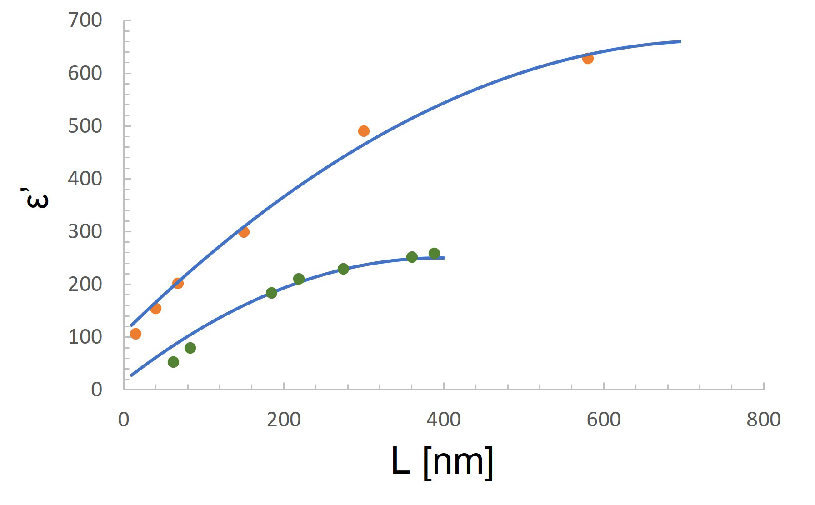}}}
}}
\caption{
         {\bf Top left}: Theoretically calculated  permittivities ``{\it $\epsilon_{\infty}$ (circles) and $\epsilon_0$ (disks) of the oxide overlayer {\it vs.} 
         its thickness, compared to the results of a classical three-layer model (solid). The classical two-layer model gives a constant permittivity corresponding
         to that of bulk SiO$_2$ (dashed). Inset: ionic contribution to $\epsilon_0$}''.
         {\bf Top right}: ``{\it High-frequency (upper panel) and static (lower panel) microscopic polarization along $x$ (solid lines) compared to the classical 
         three-layer model (dashed lines)}'' [assuming the bulk values of $\epsilon_0$ for Si and SiO$_2$]. ``{\it The polarization is normalized by its average
         value in the cell. For the static polarization we used a Gaussian broadening of 0.9~$\AA$ to smear out the pointlike ionic contributions. The suboxide
         region is shaded.}'' [Figures and words in quotes/italic reprinted with permission from F.~Giustino, P.~Umari, and A.~Pasquarello, Phys. Rev. Lett. 
         {\bf 91}, 267601 (2003), \copyright2003 by the American Physical Society.]
         {\bf Bottom left}: Experimentally measured dielectric ``{\it permittivity {\it vs}. frequency curves for different HfO$_2$ film}'' (deposited on Si by 
         sputtering and with unpassivated top surface) ``{\it thicknesses: 13.5~nm (a), 29.6~nm (b), 48.8~nm (c), and 98.6~nm (d).}''
         [Figure and words in quotes/italic reprinted with permission from D.~A.~Golosov, N.~Vilya, S.~M.~Zavadski, S.~N.~Melnikov, A.~V.~Avramchuk, 
         M.~M.~Grekhov, N.~I.~Kargin, and I.~V.~Komissarov, Thin Solid Films {\bf 690}, 137517 (2019), \copyright2019by the American Vacuum Society.] 
         {\bf Bottom right}: ``{\it Comparison between the theoretical prediction given by Eq.~(15), continuous line, and experimental data (circles). The latter are 
         in arbitrary physical units, as customary for dielectric spectroscopy measurements. The upper curve refers to experimental data of dielectric permittivity
         of (Ba$_{0.7}$, Sr$_{0.3}$)TiO$_3$ (BST) thin films measured at $\omega$ = 4~kHz from Ref.~[3], fitted by Eq.~(15) with K$_1$ = 1.5 and K$_2$ = 0.001, 
         and $\epsilon_{\infty}$ = 100. The lower curve refers to experimental data of dielectric permittivity of (Ba$_{0.5}$, Sr$_{0.5}$)TiO$_3$ thin films averaged
         between $\omega$ = 400~Hz and $\omega$ = 10~kHz from Ref.~[24], fitted by Eq.~(15) with K$_1$ = 1.2 and K$_2$ = 0.0015, and $\epsilon_{\infty}$ = 10. 
         All experimental measures were made at room temperature}.''[Figure and words in quotes/italic reprinted with permission from A. Zaccone, Phys. Rev. 
         B {\bf 109}, 115435 (2024), \copyright2024 by the American Physical Society.]} 
\label{fig:epsilon_oxides}
\end{figure*}

From these considerations, it is safe to assume that the possible dependence of the dielectric constant on the thickness of polar insulators is determined mainly
by their ionic response. A review of the literature regarding the effect of thickness and phonon confinement on the dielectric constant, $\epsilon_{\rm ox}$, of 
thin insulators shows inconsistent experimental trends and theoretical predictions. Experimentally, some studies have shown the expected reduction of 
$\epsilon_{\rm ox}$ in thin films. For example, Ref.~\cite{Golosov_2019} shows a drop of $\epsilon_{\rm ox}$ from 21.3 to 8~$\epsilon_{\rm vac}$ in HfO$_2$ films 
as their thickness is reduced from 98.6 to 13.5~nm. Similarly, Zhang {\it et al.}~\cite{Zhang_2000} have reported a drop
of the dielectric constant with a decreasing thickness of Ta$_2$O$_5$/HfO$_2$, Ta$_2$O$_5$/ZrO$_2$, and ZrO$_2$/HfO$_2$ nanolaminates. On the contrary, Perkins and
coworkers~\cite{Perkins_2001} and Chang {\it et al.}~\cite{Chang_2002} have hinted at the possible observation of a higher $\epsilon_{\rm ox}$ in thin ZrO$_2$ and
SiO$_2$ films. Theoretical calculations show the same inconsistency: For example, Giustino, Umari, and Pasquarello~\cite{Giustino_2003} have used DFT to study the 
Si-SiO$_2$ system, showing a slight increase of the SiO$_2$ dielectric constant (static and optical) in ultra-thin films ($t_{\rm ox} \lessapprox$ 0.7~nm). They
attribute this result to the presence of a 0.32~nm-thick interfacial layer with a static dielectric constant of 7.1~$\epsilon_{\rm vac}$. This is due to the
gradient of the chemical composition going from Si to SiO$_2$ via a sub-oxide SiO$_x$ ($x < 2$) transition layer and may be also be the result of the 
increased polarizabilty of the SiO$_2$ ions due to the proximity of a material, Si, with a higher polarizability, as we mentioned before. They show
that a classical three-layer model (Si/interface-sub-oxide/ SiO$_2$) fits the DFT calculation with a 2\% accuracy.\\

On the other hand, Zaccone~\cite{Zaccone_2024} 
has recently predicted a dramatic reduction of the ionic component of dielectric constant of thin films of polar materials. Whereas 
Giustino {\it et al.}~\cite{Giustino_2003} stress the importance of interface polarization effects, especially in the interfacial sub-oxide transition layer, 
Zaccone reaches his conclusion on the truncation of the Debye phonon density of states (DoS) at the lower bound $\omega_{\rm min}= c_{\rm s}\pi/t_{\rm ox}$, 
where $c_{\rm s}$ is the sound velocity and $t_{\rm ox}$ the film thickness. In general, the Debye model describes correctly the DoS of acoustic phonons.
These may affect the dielectric response only via indirect, higher-order effects, such as the elasto-optic effect~\cite{Wemple_1970} (due to changes of the charge
density in the crystal due to the strain associated with acoustic phonons), or Brillouin scattering~\cite{Bennett_1972} that may change the high-frequency response.
However, acoustic phonons play a negligible role at the low frequencies of interest here: The static dielectric constant is determined almost exclusively by the
optical phonons. Therefore, approximating the phonons DoS with the Debye model may be correct in the case of dielectrics that exhibit extremely soft optical modes,
such as barium-strontium-titanate (Ba$_x$Sr$_{1-x}$TiO$_3$, BST), on which Zaccone is interested.
However, when considering the `harder' insulators of interest here, the use of the Debye model appears inappropriate. Equally inappropriate seems to be the 
idea of setting a lower bound of the phonon frequency at $\omega_{\rm min}= c_{\rm s}\pi/t_{\rm ox}$, since the dispersion of the optical phonons is 
almost flat and does not vanish at long wavelengths. Thus, the dependence on $t_{\rm ox}$ of the static dielectric constant can only be 
due to the onset of discretized branches of confined optical phonons and Zaccone's conclusions cannot be extended 
to the insulators that are widely used in the nanoelectronic technology, such as SiO$_2$, Si$_3$N$_4$, Al$_2$O$_3$, or HfiO$_2$.\\

The results of the various articles mentioned here are summarized in Fig.~\ref{fig:epsilon_oxides}. This figure shows the contrasting experimental and theoretical results regarding the dielectric constant of thin SiO$_2$ and HfO$_2$ films and Zaccone's results for BST. Here, we intend to (hopefully ) clarify the situation by
giving a proverbial `back-of-the-envelope' estimation of the reduction of the dielectric constant that we 
may expect (or not) from the confinement of optical phonons in thin films of polar insulators. Necessarily, but regretfully, we will have to ignore interface
polarization and anisotropic effects whose treatment requires much more than the back of an envelope. \\
      
Let's start by considering bulk dielectrics, assumed isotropic (so that all vector fields defined below, ${\boldsymbol{\delta}}$, ${\bf D}$, ${\bf P}$, and 
${\boldsymbol{\rho}}_{\rm pol}$, are aligned along the same direction, in-plane or out-of-plane, of the field ${\boldsymbol{\mathcal E}}$, and we will consider 
only their magnitudes). Equation~(\ref{eq:epsox_A3}) below gives the frequency dependence of the the long-wavelength dielectric function of polar crystals.
Following roughly Zaccone's treatment of the problem~\cite{Zaccone_2024},  
this expression can be obtained by considering a simple linear diatomic chain (mimicking a polar crystal) and noticing that, in the long-wavelength limit, 
the distance $\delta$ between the anion and the cation oscillates in time according to the simple harmonic equation of motion:
\begin{equation}
\frac{{\rm d}^2 \delta(t)}{{\rm d}t^2} - \gamma \frac{{\rm d} \delta(t)}{{\rm d}t} + \omega_{\rm TO}^2 \delta(t) = 0 \ ,  
\label{eq:ionic_distance}
\end{equation}
where $\omega_{\rm TO} = (D/\mu)^{1/2}$ is the long-wavelength frequency of the transverse optical phonons, $D$ is the `spring constant' and, in the dielectric 
continuum limit, $\mu$ is the reduced mass of ions 1 and 2, $\mu^{-1}=m_1^{-1}+m_2^{-1}$. we have also included a `friction' term
$-\gamma [{\rm d}\delta(t)/{\rm d}t]$, where $\gamma=1/\tau_{\rm 3-ph}$, that mimics the finite lifetime $\tau_{\rm 3-ph}$ of the optical phonons due to their
anharmonic coupling to the acoustic phonons). In the presence of an oscillating electric field, 
$\mathcal{E}e^{i \omega t}$, the anion-cation distance takes the form $\delta(t)=\delta_0(\omega) e^{i \omega t}$ where:
\begin{equation}
\delta_0(\omega) = - \frac{e^{\ast}\mathcal{E}}{\mu} \frac{1}{\omega_{\rm TO}^2-\omega^2 -i \omega \gamma} \ ,
\label{eq:ionic_resonance}
\end{equation}
where $e^{\ast}$ is the effective charge of each dipole. From now on, we will ignore the `broadening' $\gamma$, since it does not affect 
the main point of this discussion. For a volume density $n_{\rm dip}$ of anion-cation pairs, the polarization density (in C/m$^2$) will be:
\begin{equation}
\rho_{\rm pol}(\omega) =  n_{\rm dip} e^{\ast} \delta_0(\omega) =  
              - n_{\rm dip} \frac{e^{\ast 2}\mathcal{E}}{\mu} \frac{1}{\omega_{\rm TO}^2-\omega^2} \ .
\label{eq:polarization_charge}
\end{equation}
Finally, the polarization field $P(\omega)$ will be (assuming a uniform $\epsilon^{(\infty)}$, the electronic component of the dielectric response):
\begin{equation}
P(\omega) = - \frac{\rho_{\rm pol}(\omega)} {\epsilon^{(\infty)}} =  
          \frac{n_{\rm dip} e^{\ast 2}\mathcal{E}}{\epsilon^{(\infty)} \mu} \frac{1}{\omega_{\rm TO}^2-\omega^2} \ .
\label{eq:polarization_field}
\end{equation}
where $\epsilon^{(\infty)}$ is the dielectric constant at a frequency large enough for the phonons not to respond. In practice this will be the static, 
long-wavelength dielectric constant due exclusively to the electronic response. As a result, the displacement field $D$ in this simple model will be
\begin{equation}
D(\omega) = \epsilon^{(\infty)} \mathcal{E} + \epsilon^{(\infty)} P(\omega) = 
          \epsilon^{(\infty)} \mathcal{E} \left [ 1 + \frac{n_{\rm dip} e^{\ast 2}}{\epsilon^{(\infty)} \mu} \frac{1}{\omega_{\rm TO}^2-\omega^2}\right ]  \ .
\label{eq:polarization_field_2}
\end{equation} 
In this equation we can identify the dielectric function $\epsilon_{\rm ox}(\omega)$ as
\begin{equation}
\epsilon_{\rm ox}(\omega) = 
          \epsilon^{(\infty)} + \frac{n_{\rm dip} e^{\ast 2}}{\mu} \frac{1}{\omega_{\rm TO}^2-\omega^2} 
          = \epsilon^{(\infty)} + f \frac{\omega_{\rm TO}^2}{\omega_{\rm TO}^2-\omega^2} \ .
\label{eq:epsox_A1}
\end{equation}
The factor $f = n_{\rm dip} e^{\ast 2}/(\mu \omega_{\rm TO}^2)$ is proportional to the oscillator strength of the phonons. It determines how the phonon response 
increases the dielectric response from $\epsilon^{(\infty)}$, at frequencies large enough so that the phonons do not respond, to a `static' value
$\epsilon^{(0)}$ affected by the full ionic response. Therefore, we may recast Eq.~(\ref{eq:epsox_A1}) as:
\begin{equation}
\epsilon_{\rm ox}(\omega) = 
          \epsilon^{(\infty)} + (\epsilon^{(0)}-\epsilon^{(\infty)}) \frac{\omega_{\rm TO}^2}{\omega_{\rm TO}^2-\omega^2}  \ .
\label{eq:epsox_A2}
\end{equation}
Considering the more realistic case of a optical phonons with a dispersion $\omega^{\prime}({\bf q})$, then Eq.~(\ref{eq:epsox_A1}) becomes:
\begin{equation}
\epsilon_{\rm ox}(\omega) = \epsilon^{(\infty)} + 
        \int {\rm d} \omega^{\prime} \ f(\omega^{\prime}) {\mathcal{D}}^{\rm (TO)}(\omega^{\prime}) \frac{\omega^{\prime 2}}{\omega^{\prime 2}-\omega^2}  \ ,
\label{eq:epsox_A1b}
\end{equation}
where ${\mathcal{D}}^{\rm (TO)}(\omega)$ is the DoS of the TO phonons at frequency $\omega$ and $f(\omega)$ their oscillator strength. 
This expression can be simplified by noticing that in most insulators the optical 
phonons exhibit a small dispersion. For example, because of the relatively small mass difference between Si and O, in SiO$_2$ 
${\mathcal{D}}^{\rm (TO)}(\omega)$ exhibits two sharp phonon-DoS peaks at approximately 45-50 and 140-150~meV in both the $\alpha$-quartz and $\beta$-cristobalite forms. 
Even though HfO$_2$ may exhibit a stronger dispersion, it is reasonable to ignore the TO dispersion altogether and consider an 
approximate TO DoS consisting of two Dirac-deltas at two frequencies, $\omega_{\rm TO1}>\omega_{\rm TO2}$ (we shall convert the integral in Eq.~(\ref{eq:epsox_A1b}) 
to a sum over discrete modes also in Eq.~(\ref{eq:epsox_A4}) below, when dealing with confined phonons.) Therefore, Eq.~(\ref{eq:epsox_A1b}) can be approximated as:
\begin{multline}
\epsilon_{\rm ox}(\omega) = 
          \epsilon^{(\infty)} + (\epsilon^{\rm (mid)}-\epsilon^{(\infty)}) \frac{\omega_{\rm TO1}^2}{\omega_{\rm TO1}^2-\omega^2} + \\
             \ (\epsilon^{(0)}-\epsilon^{\rm (mid)}) \frac{\omega_{\rm TO2}^2}{\omega_{\rm TO2}^2-\omega^2} \ ,
\label{eq:epsox_A3}
\end{multline}
where $\epsilon^{\rm (mid)}-\epsilon^{(\infty)}= f_1$ and $\epsilon^{(0)}-\epsilon^{\rm (mid)}=f_2$ are the oscillator strengths of the two TOs.\\

What we wish to consider now is how Eq.~(\ref{eq:epsox_A3}) changes in thin insulating films. In principle, we expect that in these films the phonon DoS,
${\mathcal{D}}^{\rm (TO)}(\omega)$, will change, since the TO phonons are geometrically confined. Indeed, in a bulk insulator, the total oscillator strength
arises from the contribution of all the TO bulk modes. So, for example, if we are interested in looking at the long-wavelength ionic response on the $(x,y)$ plane of
the crystal, TO phonons with any out-of-plane wavevector $q_z$ will contribute to $f$. The dielectric function $\epsilon_{\rm ox}$ will be obtained from
Eq.~(\ref{eq:epsox_A3}) with a spatially constant dipole density $n_{\rm dip}e^{\ast}\delta_0$, as discussed above. However, in a thin film  of thickness 
$t_{\rm ox}$ lying on the $(x,y)$ plane, there will be only a discrete number of modes confined along the out-of-plane $z$ direction that can contribute to the
polarizability. The problem now consists in determining the fractional oscillator strength of each one of these modes.\\

Let's proceed as follows: Let's assume that the ionic displacement $\delta$ is clamped to zero at the surfaces/interfaces at $z=0$ and $z=t_{\rm ox}$. 
This is the extreme worst-case scenario, since it will result in the lowest phonon density of states and so, in the lowest dielectric constant. In this case,  
for each confined TO phonon we will have a spatially-varying anion-cation separation $\delta(z)$ proportional to $\sin(q_{z,n}z)$ where 
$q_{z,n} = (2n+1) \pi/t_{\rm ox}$ is the discretized wavevector of the TO phonon. Since we are interested in estimating the dielectric constant of 
the film averaged over its thickness, only symmetric modes ({\it i.e.}, symmetric around the center of the film, corresponding to odd indices $2n+1$) will 
give a non-vanishing contribution to the thickness-averaged polarization; therefore, we can ignore the antisymmetric/odd modes. Then,
keeping in mind that, for a spatially varying polarization density ${\boldsymbol{\rho}}_{\rm pol}$, the polarization field ${\bf P}$ is given by
$\nabla \cdot {\bf P} = - \rho_{\rm b}/\epsilon^{(\infty)} = - \nabla \cdot {\boldsymbol{\rho}}_{\rm pol}/\epsilon^{(\infty)}$
(where $\rho_{\rm b}=\nabla \cdot {\boldsymbol{\rho}}_{\rm pol}$ is the bound charge density),
the ionic polarizability will result in a position-dependent long-wavelength local dielectric function $\epsilon_{\rm ox}(\omega, z)$: 
\begin{multline}
\epsilon_{\rm ox}(\omega, z) = \epsilon^{(\infty)} + S_{N}(z) \left [
(\epsilon^{\rm (mid)}-\epsilon^{(\infty)}) \frac{\omega_{\rm TO1}^2}{\omega_{\rm TO1}^2-\omega^2} + \right. \\ \left.
             \ (\epsilon^{(0)}-\epsilon^{\rm (mid)}) \frac{\omega_{\rm TO2}^2}{\omega_{\rm TO2}^2-\omega^2} \right ] \ ,
\label{eq:epsox_A4}
\end{multline}
where the function $S_{N_{\rm s}}(z)$ that modulates the phonon oscillator strength is
\begin{equation}
S_{N_{\rm s}}(z) = 
\left \{ \begin{array}{ll}
\frac{4}{\pi} \sum_{n=0}^{N_{\rm s}} \frac{1}{2n+1} \sin \left [ \frac{(2n+1) \pi}{t_{\rm ox}} z \right ] & \mbox{   for } 0 \le z < t_{\rm ox} \\
                                              0                                                           & \mbox{   otherwise} \\ 
         \end{array} \right. \ ,
\label{eq:epsox_A5}
\end{equation}
where the fractional oscillator strength (or 'weight') of each confined phonon is $4/[\pi(2n+1)]$. This can be seen in two different (but mathematically equivalent)
ways: As explained in detail in footnote~\cite{enote5}, this weight is the areal density of in-plane TO-phonon states for a given $q_{z,2n+1}$, normalized to the total
density of modes. Alternatively, note that Eq.~(\ref{eq:epsox_A5}) is the simple Fourier-series expansion of a square wave. Therefore, Eq.~(\ref{eq:epsox_A5}) 
yields a constant value (unity) for $0 \le z < t_{\rm ox}$. This gives the expected $z$-independent ionic polarization when summing over the infinitely many modes
present in an infinitely thick film. In summary, Eq.~(\ref{eq:epsox_A5}) reflects the $z$-dependent dipole density, $n_{\rm dip} e^{\ast} \delta_0 S_{N_{\rm s}}(z)$,
of confined TO phonons and yields the expected position-independent dielectric constant in an infinitely thick film (that is, in the limit of a bulk dielectric
as $N_{\rm s} \rightarrow \infty$).\\ 
\begin{figure*}[tb!]
\centerline 
{\hbox{
\includegraphics[width=18.00cm]{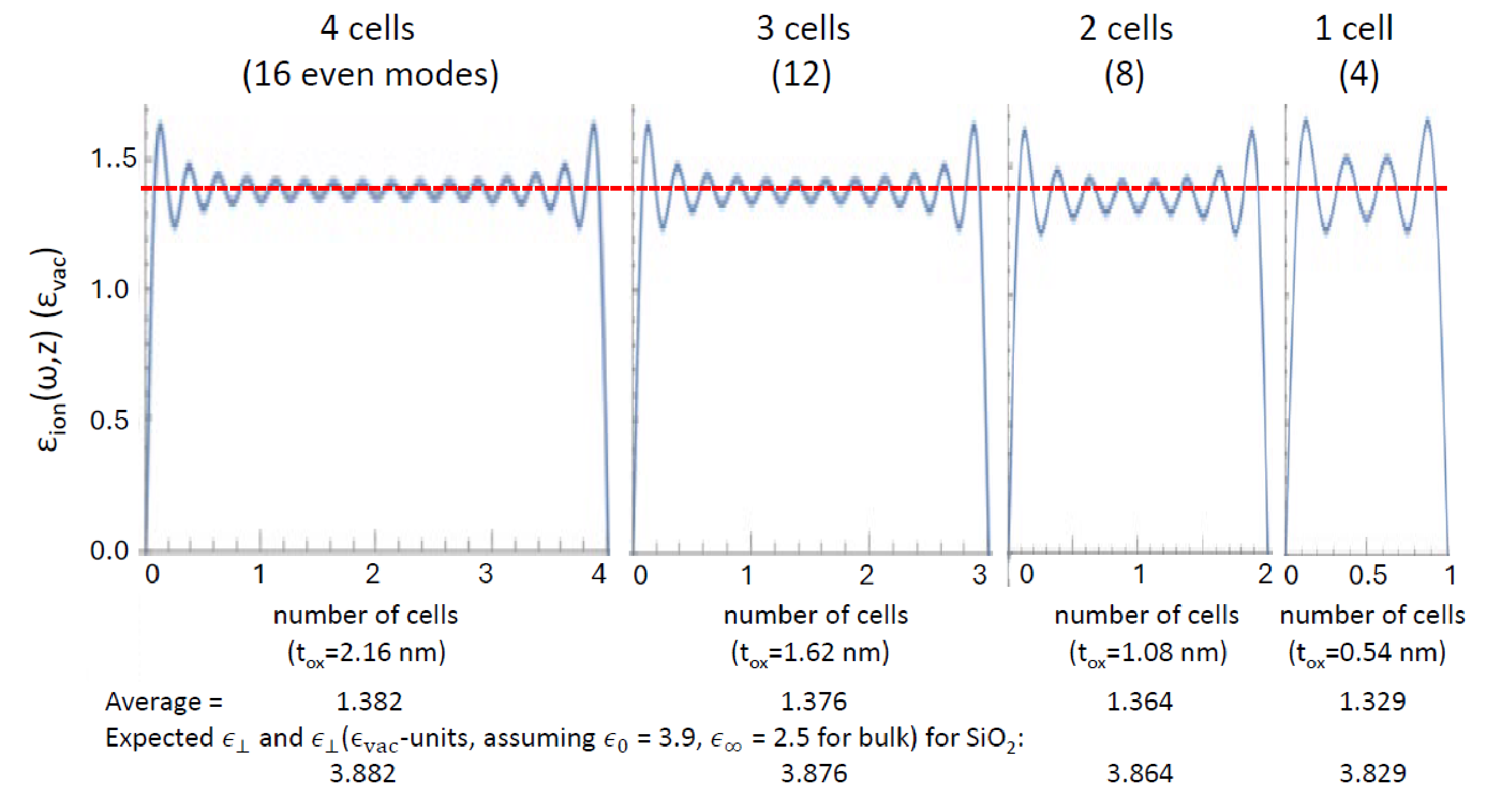}}}
\caption{Spatial dependence of the ionic component of the SiO$_2$ dielectric constant) expressed by Eqs.~(\ref{eq:epsox_A4}) and (\ref{eq:epsox_A5}) 
         with the sum extended over the  number of even TO modes in a (100) SiO$_2$ film with a thickness of 4, 3, 2, and 1 cells (from left to right). 
         Although fewer modes contribute in thinner films, the average dielectric constant, as shown assuming a  thickness-independent electronic response, 
         $\epsilon^{(\infty)}$), decreases only very slightly. The peaks seen at the surfaces of the layer are artifacts due to the Gibbs phenomenon. 
         They may distort the results to some extent.}  
\label{fig:Oscillator-strength}
\end{figure*}

Obviously, the number of symmetric (even) modes, $N_{\rm s}$, appearing in the sum in Eq.~(\ref{eq:epsox_A5}) is determined by the number of even TO-phonon modes  present in the structure. This is $N/2-1$, where $N$ is the total number of TO phonons. For a layer $N_{\rm c}$-cell thick with $N_{\rm a}$ atoms in each cell, 
we have $N = N_{\rm c}(N_{\rm a}-1)$ out-of-plane and $N_{\rm c}(N_{\rm a}-1)$ in-plane TO phonons. The oxides we are interested in are likely to be amorphous.
However, considering them to be in some
crystalline structure to get an estimate of the modes we should consider, $N_{\rm a}=9$ for SiO$_2$ in the $\alpha$-quartz structure and $N_{\rm a}=24$ in the 
$\beta$-cristobalite structure. For HfO$_2$, $N_{\rm a}=12$
in the orthorhombic, cubic, and monoclinic crystal structures, whereas $N_{\rm a}=6$ in the tetragonal form. Therefore, in the limit of an infinitely thick film
($N_{\rm c} \rightarrow \infty$), $S_{N_{\rm s}}(z) \rightarrow S_{\infty}(z) = \theta(z) - \theta(z-t_{\rm ox})$, the `square-wave' function that is equal to 1 
for $0 \le z < t_{\rm ox}$, zero otherwise. In this limit, $\epsilon_{\rm ox}(\omega,z)$ is the `usual' dielectric function,
spatially constant in the film. However, considering thin SiO$_2$ films in the $\alpha$-quartz structure (just to fix the ideas for an interesting case),
the number of modes will be relatively small: $N_{\rm s}=3$ (4 even modes) for a 1-cell ($\approx 0.541$~nm) thick film; 
$N_{\rm s}=7$ (8 even modes) for a 2-cell ($\approx 1.08$~nm) thick film; $N_{\rm s}=11$ (12 even modes) for a 3-cell ($\approx 1.62$~nm) thick film.\\

Figure~\ref{fig:Oscillator-strength} shows the $z$-dependent ionic component of the local dielectric `constant' 
$\epsilon_{\rm ion}(\omega=0,z)=[\epsilon^{(0)}-\epsilon^{(\infty)}] S_{N_{\rm s}}(z)/\epsilon_{\rm vac}$ for SiO$_2$ films of thickness decreasing from 
4 $\alpha$-quartz cells ($\approx$ 2.16~nm) to 1 cell ($\approx$ 0.54~nm), assuming $[\epsilon^{(0)}-\epsilon^{(\infty)}]\epsilon_{\rm vac}=1.4$, the bulk value.
Although the dielectric constant vanishes at the surfaces of the films, an obvious consequence the clamped boundary conditions for the TO phonons, its average
value remains very close to the bulk value: the polarizability is almost completely determined by the mode with the longest wavelength along the $z$
direction (that is, the $2n+1$ mode with $n=0$). This is because $8/\pi^2 \approx 81$\% of the bulk ionic polarizability 
is due to this mode, thanks to the large number of in-plane states with such a small out-plane-wavevector; as films become thinner and high-$2n+1$ modes are 
`lost', the effect of the phonon confinement is minor. As mentioned above, 
this is the worst-case scenario: If the TOs are free-standing or confined over a thicker region, the dielectric constant would be even closer to the bulk value. Moreover, the `peaks observed near the surfaces in
Fig.~\ref{fig:Oscillator-strength} are likely to be an artifact due to the Gibbs phenomenon. The net result of this `back-of-the-envelope' analysis is that 
the confinement of the TO phonons in thin insulators should not affect the dielectric constant in any significant way.\\

\section{\label{sec:conclusion}{Conclusions}}
We have argued that the static dielectric constant of thin semiconductor and insulator nanostructures is affected by the materials in which they are 
embedded. As fas as the electronic responses is concerned, we have reached this conclusion simply by reviewing published experimental and theoretical 
results. On the contrary, regarding the ionic response, we have used a simple 'worst-case scenario' model to consider the reduced density of states of
optical phonons in thin films to show that this has a negligible effect on their static dielectric constant. As a practical consequence of our study,
we have argued hat when studying the electrostatic and charge-transport properties of electronic devices scaled at to sub-10~nm length, the use of the
bulk dielectric constant is a satisfactory approximation. \\ 

\acknowledgments
This work has been supported by the Taiwan Semiconductor Manufacturing Company Ltd. (TSMC). We also thank the Texas Advanced Computing Center (TACC) for having provided the computing resources required to perform this study and Texas Instruments for the Endowment that has provided additional financial support.

\bibliography{paper_ref_titles}

\end{document}